%% file: main.tex
\documentclass[11pt,a4paper,reqno]{amsart}

\usepackage{times}
\usepackage{mathrsfs,mathtools}
\usepackage{enumerate}
\usepackage{stmaryrd} 
\usepackage{amsmath,amsthm,amssymb,amsfonts, accents} 
\usepackage[usenames,dvipsnames]{xcolor}
\usepackage{subfiles}
\usepackage{dsfont}
\usepackage[mathscr]{euscript}
\usepackage{graphicx}
\usepackage[colorlinks=true, linkcolor=tub-red,citecolor=my-blue]{hyperref}
\usepackage[T1]{fontenc}
\usepackage[english]{babel}
\usepackage{todonotes}
\usepackage{geometry}
\usepackage{xspace,cancel}
\usepackage[capitalize]{cleveref}
\crefname{subsection}{Subsection}{Subsections}
\usepackage[numbers,sort&compress,longnamesfirst]{natbib}
\usepackage{lipsum}  
\usepackage{extarrows}
\usepackage{relsize}
\usepackage{bbm}
\usepackage{blkarray} 
\usepackage{pgf,tikz}
\usepackage{mathrsfs}
\usepackage[capposition=bottom]{floatrow}
\usepackage{relsize}
\usepackage{mathtools}
\usepackage{fancyhdr}
\usepackage{bm}
\usepackage{breqn}

\def\d{\mathrm{d}}

\definecolor{my-blue}{RGB}{0,85,150}
\definecolor{tub-red}{RGB}{187,14,31}

\makeatletter \@addtoreset{equation}{section}

\newtheorem{theorem}{Theorem}[section]
\newtheorem*{assumption*}{\assumptionName}
    \providecommand{\assumptionName}{}
    

  \makeatother

\theoremstyle{definition}

\newtheorem{remark}[theorem]{Remark}

\makeatletter
\def\namedlabel#1#2{\begingroup
    #2%
    \def\@currentlabel{#2}%
    \phantomsection\label{#1}\endgroup
}

\numberwithin{equation}{section}
\numberwithin{figure}{section}
\numberwithin{table}{section}

\newcommand{\tnorm}[1]{\left\vert\kern-0.25ex\left\vert\kern-0.25ex\left\vert #1 \right\vert\kern-0.25ex\right\vert\kern-0.25ex\right\vert}
\newcommand{\tnormbig}[1]{\bigl\vert\kern-0.25ex\bigl\vert\kern-0.25ex\bigl\vert #1 \bigr\vert\kern-0.25ex\bigr\vert\kern-0.25ex\bigr\vert}

\newcommand{\ud}{\ensuremath{\mathrm{d}}}

\newcommand{\dz}{\ud z}
\newcommand{\dy}{\ud y}
\newcommand{\e}{\mathrm{e}}

\newcommand{\levy}{L\'evy\xspace}

\DeclareFontFamily{U}{mathx}{\hyphenchar\font45}
\DeclareFontShape{U}{mathx}{m}{n}{
<5><6><7><8><9><10>
<10.95><12><14.4><17.28><20.74><24.88>
mathx10
}{}
\DeclareSymbolFont{mathx}{U}{mathx}{m}{n}
\DeclareFontSubstitution{U}{mathx}{m}{n}
\DeclareMathAccent{\widebar}{0}{mathx}{"73}

\begin{document}

\title[A Deep IMEX Minimizing Movement Method for Option Pricing in Jump-Diffusion Models]{A Deep Implicit-Explicit Minimizing Movement Method \\ for Option Pricing in  Jump-Diffusion Models}

\author[E.~H. Georgoulis]{Emmanuil H. Georgoulis}
\author[A. Papapantoleon]{Antonis Papapantoleon}
\author[C. Smaragdakis]{Costas Smaragdakis}

\address{Department of Mathematics and The Maxwell Institute for Mathematical Sciences, Heriot-Watt University, Edinburgh EH14 4AS, United Kingdom  -- \& --  Department of Mathematics, School of Applied Mathematical and Physical Sciences, National Technical University of Athens, 15780 Zografou, Greece -- \& --  Institute of Applied and Computational Mathematics, FORTH, 70013 Heraklion, Greece}\email{e.georgoulis@hw.ac.uk}

\address{Delft Institute of Applied Mathematics, TU Delft, 2628 Delft, The Netherlands -- \& --  Department of Mathematics, School of Applied Mathematical and Physical Sciences, National Technical University of Athens, 15780 Zografou, Greece -- \& --  Institute of Applied and Computational Mathematics, FORTH, 70013 Heraklion, Greece}
\email{a.papapantoleon@tudelft.nl}

\address{Department of Statistics and Actuarial-Financial Mathematics, University of the Aegean, 83200 Karlovassi, Samos, Greece  -- \& --  Institute of Applied and Computational Mathematics, FORTH, 70013 Heraklion, Greece}
\email{kesmarag@aegean.gr}



\thanks{This research work was supported by the Hellenic Foundation for Research and Innovation (H.F.R.I.) under the ``First Call for H.F.R.I. Research Projects to support Faculty members and Researchers and the procurement of high-cost research equipment grant'' (Project Number: 2152).
The authors also acknowledge the use of computational resources of the DelftBlue supercomputer, provided by the Delft High-Performance Computing Centre \cite{DHPC2022}. 
EHG also wishes to acknowledge the financial support of The Leverhulme Trust (grant number RPG-2021-238) and of EPSRC (grant number EP/W005840/2).}

\date{}

\begin{abstract}
We develop a novel deep learning approach for pricing European basket options written on assets that follow jump-diffusion dynamics. 
The option pricing problem is formulated as a partial integro-differential equation, which is approximated via a new implicit-explicit minimizing movement time-stepping approach, involving approximation by deep, residual-type Artificial Neural Networks (ANNs) for each time step. 
The integral operator is discretized via two different approaches: (a) a sparse-grid Gauss--Hermite approximation following localised coordinate axes arising from singular value decompositions, and (b) an ANN-based high-dimensional special-purpose quadrature rule. 
Crucially, the proposed ANN is constructed to ensure the appropriate asymptotic behavior of the solution for large values of the underlyings and also leads to consistent outputs with respect to \emph{a priori} known qualitative properties of the solution.
The performance and robustness with respect to the dimension of these methods are assessed in a series of numerical experiments involving the Merton jump-diffusion model, while a comparison with the deep Galerkin method and the deep BSDE solver with jumps further supports the merits of the proposed approach.
\end{abstract}

\maketitle

\frenchspacing


\section{Introduction} 
\subfile{introduction.tex}

\section{Option pricing in jump-diffusion models}
\label{model_problem} 
\subfile{deepRitz.tex}

\section{Deep implicit-explicit minimizing movements and network architecture} 
\label{imex}
\subfile{modeling.tex}

\subsection{Computation of the integral operator} 
\subfile{integral.tex}

\section{Numerical examples}  
\label{examples}
\subfile{applications.tex}

\section{Conclusions}
\label{conclusions}
\subfile{conclusions}

\bibliographystyle{abbrvnat}
\bibliography{references}

\end{document}

%% file: introduction.tex
A central problem in Mathematical Finance is the fast and accurate computation of arbitrage-free prices of financial derivatives, especially for advanced stochastic models and for multi-asset derivatives.
A basket option is a contractual agreement between two parties, the buyer and the seller, to buy or sell a derivative whose value fluctuates over time based on the prices of a set of underlying assets (the ``basket''). 
In this work, we consider the problem of pricing European basket call options over \(d\) underlying assets using weights \(\{\alpha_i\}_{i=1}^d\). 
The strike price, denoted by \(K\), is the price at which the basket can be bought at the maturity of the option contract, \textit{i.e.} at \(t=T\).
The payoff of a European basket call option is provided by
\[
    \biggl(\sum_{i=1}^d \alpha_i S^i_T - K \biggr)^+,
\]
where \(\alpha_i>0\) and \(\sum_{i=1}^d \alpha_i = 1\). 
The value \(S^i\) of each asset is associated to the variable \(x_i =S_T^i/K\), referred to in the literature as the moneyness of the stock.
The payoff function of a basket call option is then provided by
\begin{equation}\label{payoff_function_ic}
\mathrm{Payoff}(x) = u_0(x) := \biggl(\sum_{i=1}^d \alpha_i x_i - 1 \biggr)^+.
\end{equation}

Models for asset prices that incorporate random jumps are essential in capturing the real-world behavior of financial markets. 
These models acknowledge that financial asset prices do not always move smoothly; instead, they may present abrupt changes due to unexpected events.
Incorporating random jumps into asset price models helps to obtain a better description of the real-world behavior of the markets.
Indeed, they allow to capture the fat-tails and skews present in asset log-returns under the ``real-world'' measure, while they exhibit a volatility smile or skew under the ``risk-neutral'' measure.
We refer the reader to \citet{Eberlein_Kallsen_2019, Cont_Tankov_2004} or \citet{Schoutens_2003} for more details and references on models with jumps in finance.

A popular model that incorporates random jumps is the, so-called, Merton \cite{Merton_1976} model, which belongs to the family of \levy jump-diffusion models, \textit{i.e.} jump-diffusion models with stationary and independent increments. 
In the Merton model, asset prices follow a superposition of a geometric Brownian motion and a Poisson process with randomly distributed jumps, allowing for both continuous diffusion, as in the Black--Scholes model, and discrete random jump movements, realised by a multivariate normal distribution for the jump size. 
The intensity and the magnitude of the jumps are determined by parameters that can be estimated from market data, \textit{e.g.} option prices of single- and multi-asset options.
Other popular jump-diffusion models in the literature are the \citet{Kou_2002} model and affine jump-diffusions, see \textit{e.g.} \citet{Bates_1996} or \citet{Duffie_Pan_Singleton_2002}.

{ The industry standard for pricing European single-asset options in \levy and affine jump-diffusion models are transform methods, such as Fourier or COS; see \textit{e.g.} \citet{Carr_Madan_1999,Fang_Oosterlee} or \citet{Eberlein_Glau_Papapantoleon_2010}.
These methods, however, suffer from the curse of dimensionality and can typically not be applied for the valuation of basket options in dimensions higher than 8 or 10; see \citet{Bayer_BenHammouda_Papapantoleon_Samet_Tempone_2023,Bayer_BenHammouda_Papapantoleon_Samet_Tempone_2024} for the state of the art in this direction.}

An alternative and general method for pricing options in \levy and affine models is to solve the associated second-order partial integro-differential equation (PIDE), where the initial (terminal) condition is determined by the payoff function \eqref{payoff_function_ic} of the option.
The PIDE is derived by the Fundamental Theorem of Asset Pricing and the Feynman--Kac lemma, which relates the discounted expectation of the payoff with an initial (terminal) value problem of the form described in the following section. 
{ Once again, finite difference and finite element discretizations for the solution of the option pricing PIDEs suffer from the curse of dimensionality and cannot be used in dimensions higher than 6 or 8; see \textit{e.g.} \citet{Griebel_Hullmann_2013,Hepperger_2010} and \citet{Reichmann_Schwab_2010}.}

In this work, we address the critical challenge of computing the fair prices of financial derivatives by discretizing the PIDE using a novel implicit-explicit minimizing movement approach involving Artificial Neural Networks (ANNs). 
The use of ANNs aims to address the curse of dimensionality typically encountered in standard grid-based methods, while simultaneously offering improved performance compared to standard or Quasi Monte Carlo approaches. 

The classical minimizing movement method of De Giorgi, see \citet{degiorgi}, provides discretization of gradient flows in the calculus of variations, by considering a suitable minimization functional for each time step. 
A canonical example is the minimizing movement method for the Dirichlet energy corresponding to the backward Euler timestepping for the heat equation. 
The celebrated work of Jordan, Kinderlehrer, and Otto \cite{JKO} employed the minimizing movement approach to represent Fokker--Planck equations as gradient flows of the Boltzmann entropy in the Wasserstein metric. 

In this work, we develop a significant extension of the ``deep minimizing movement'' method of \citet{Georgoulis2023,park2023} to approximate the PIDE problem arising from basket option pricing in jump-diffusion models. 
In \cite{Georgoulis2023,park2023}, parabolic problems are discretized by ANNs (in space) in a time-stepping fashion by minimizing appropriate energy cost functionals; see also \citet{Weinan2018,Liao2019} for energy minimization for elliptic PDEs. 
The resulting method in \cite{Georgoulis2023,park2023} amounts to the backward Euler timestepping (discrete gradient flow) for the Dirichlet energy minimization via a, so-called, deep Nitsche approach. 
In the current framework, the spatial operator in the PIDE model problem involves skew-symmetric and integral terms, which are (effectively) discretized in an explicit manner. 
To that end, we consider the family of implicit-explicit versions of Backward Differentiation Formulae (BDF) methods, the first order of which is the implicit-explicit Euler scheme. 
This choice is made due to the ``$A(\alpha)$-stability'' properties of BDF methods, in conjunction with the favourable computational cost in this context, since the expensive (due to the ANN architecture) spatial operator is evaluated only once for each time-step. 
We note that the minimizing movement approach from \cite{Georgoulis2023,park2023} has the important advantage of using the ANN parameters from the previous time-step as starting point in the optimization (learning) of the approximate solution parameters at the current step. 
This reduces the computational cost of optimization significantly; see \cite{Georgoulis2023}.

A further computational challenge stems from the complexity of the integral operator due to the jumps in the asset price process. 
To that end, we devise and compare two different approaches to address the curse of dimensionality there: (a) a sparse-grid Gauss--Hermite approximation following localised coordinate axes arising from singular value decompositions, and (b) an ANN-based high-dimensional special-purpose quadrature rule.

Moreover, in order to improve the overall performance and accuracy of the method, we introduce a decomposition of the solution into two parts: the option price is expressed as the sum of a non-negative unknown component \(w(t,x)\), termed in the literature the time value of the option,  and a known lower-bound function \(v(t,x)\), termed in the literature the intrinsic value of the option. 

Finally, a domain truncation method is introduced to enhance the accuracy of the numerical schemes. This involves projecting the option prices onto a bounded subset of \(\mathbb R_+^d\) and efficiently approximating the solution for extreme moneyness values by exploiting estimates within this bounded domain.

The above numerical methodology is tested through a comprehensive series of numerical experiments, showing the competitiveness of the approach against Quasi Monte Carlo in both low and high dimensional settings.

In addition, we provide a basic comparison of the proposed method against the popular deep Galerkin method (DGM) by \citet{Sirignano2018} and the deep BSDE solvers with jumps by \citet{Han_Jentzen_E_2018} and by \citet{Gnoatto2022},  showcasing the competitiveness of the proposed method.

The remainder of this work is structured as follows. 
In Section \ref{model_problem}, we introduce the PIDE for option pricing in jump-diffusion models, present the implicit-explicit minimizing movement method and discuss the decomposition of the solution into a lower bound and an unknown function, as well as the truncation of the domain and the extension of the solution to extreme cases.
In Section \ref{imex}, we describe the architecture of the neural network and the training procedure, and then present two methods for the efficient computation of the integral operator.
In Section \ref{examples}, we test the performance of the methods using the Merton model, in scenarios with 5 and 15 underlying assets.
Finally, Section \ref{conclusions} concludes this work.

%% file: deepRitz.tex
\subsection{An illustrative example: the Merton model}

Let $(S^1,\dots,S^d)$ denote \(d\) financial assets, where each one follows the, so-called, Merton model, see \citet{Merton_1976}, whereby the dynamics of the \(i\)-th stock at time \(t\) are provided by
\begin{equation}
S_t^{(i)} = S_0^{(i)} \exp\biggl( b_i t + \sigma_i W_t^{(i)} + \sum_{k=1}^{N_t}Z_{k}^{(i)}\biggr),\quad t \in \mathbb T = (0,T],\ i \in \mathbb I=\{1,\dots,d\},
\end{equation}
where \(W^{(i)}\) denotes a standard Brownian motion, $\sigma_i>0$ denotes the diffusion volatility, \(N\) denotes a Poisson process with parameter \(\lambda>0\), $T>0$ is a finite time horizon, and \(Z_k^{(i)}\) are random variables controlling the jump sizes. 
The random vector \((Z_k^{(1)},\dots,Z_k^{(d)})\) follows a multivariate normal distribution with mean \(\mu_J\) and variance-covariance matrix \(\Sigma_J\).
Moreover, assuming that we are already working under an equivalent martingale measure, the drift parameter $b_i$ is determined by the martingale condition, and equals 
\begin{align}\label{eq:martingale}
    b_i = r - \frac12 \sigma_i^2 - \lambda \Big( \exp\Big\{ \mu_{J_i}+\frac12 \sigma_{J_i}^2 \Big\}-1 \Big),    
\end{align}
where $r$ denotes the risk-free interest rate.

There are three mechanisms that generate dependence between the assets in this example. 
The assets can have correlated diffusion terms, as well as correlated jump sizes. 
Moreover, the assets share the timing of the jumps, as the jumps occur according to a \textit{common} Poisson process. 
These dependencies have a direct impact on the arbitrage-free price of an option. 
Indeed, according to the Fundamental Theorem of Asset Pricing (FTAP) and using the Feynman--Kac formula, the arbitrage-free price of an option with payoff given by \eqref{payoff_function_ic} satisfies the following PIDE:
\begin{equation}\label{merton_PIDE}
\begin{aligned}
\frac{\partial u}{\partial t} - \frac{1}{2} \sum_{i,j=1}^d \sigma_i\rho_{ij}\sigma_jx_ix_j\frac{\partial^2u}{\partial x_i\partial x_j}
 + \sum_{i=1}^d b_ix_i\frac{\partial u}{\partial x_i} + ru   - I_\varphi[u] &=0 \\
u(0,x) &=\ u_0(x),               
\end{aligned}
\end{equation}
for $t\in\mathbb T$ and $x\in[0,\infty)^d$, where $\rho_{ij}$ denotes the diffusion correlation between assets $i$ and $j$.
The integral operator is provided by
\begin{equation}
    I_\varphi[u] = \lambda \int_{\mathbb R^d} \big( u(t,x\e^z)-u(t,x) \big) \varphi(\dz),
\end{equation}
with \(\varphi(\cdot)\) denoting the probability density function of the multivariate normal distribution with mean $\mu_J$ and variance-covariance matrix $\Sigma_J$. 
Obviously, \(u(t,0)=0\) and \(I[u(t,0)]=0\), for each \(t>0\). 
The initial condition in \eqref{merton_PIDE} is derived from the terminal condition of the classical PIDE for option pricing, \textit{i.e.} from the payoff function in \eqref{payoff_function_ic}, by employing the change of variable \(t = T- \cdot\).

\subsection{General form of the PIDE for option pricing}

Motivated by \eqref{merton_PIDE}, we consider the following general problem: find the arbitrage-free price $u$ of an option with payoff \eqref{payoff_function_ic}, that satisfies the PIDE
\begin{equation}
\begin{aligned}    \label{eq:PIDE}
\frac{\partial}{\partial t} u(t,x) + \mathcal{A}u(t,x) &=0, \quad (t,x)\in \mathbb T \times [0,\infty)^d\\ 
u(0,x)&= u_0(x), \quad x \in [0,\infty)^d,
\end{aligned}
\end{equation}
where the operator \(\mathcal A\) belongs to the following family:
\begin{equation}
\mathcal{A} u(t,x) = -\sum_{i,j=1}^d a_{ij}(x)\frac{\partial^2 u}{\partial x_i \partial x_j} + \sum_{i=1}^d b_i(x)\frac{\partial u}{\partial x_i} + ru - I_\nu[u],
\end{equation}
where \(a_{ij}(x) = a_{ji}(x)\), for \(i,j\in\mathbb I\), and 
\begin{equation}
    I_\nu[u] = \lambda \int_{\mathbb R^d} \big( u(t,x\e^z)-u(t,x) \big)\nu(x,\dz),
\end{equation}
with the integration performed over a finite jump measure $\nu$.
The coefficients \(a_{ij}\) and $\nu$ are to be derived from the diffusion and the jump terms of the stochastic processes governing the asset prices, respectively, while the drift term will be determined again by the martingale condition; see \eqref{eq:martingale}. 
This class of models includes \levy jump-diffusions, such as the \citet{Kou_2002} model, and affine jump-diffusions, see \textit{e.g.} \citet{Bates_1996} or \citet{Duffie_Pan_Singleton_2002}; see also \citet{Runggaldier_2003} for a general overview of jump-diffusion models in finance.
In what follows, it is sufficient to assume that $a_{ij},b_j\in W^{1,\infty}([0,\infty)^d)$, using standard Sobolev space notation.
The initial condition in \eqref{eq:PIDE} has the meaning that the price of an option with maturity time \(T=0\) (present) is determined by its payoff function. (This, again, is the result of employing the change of variable \(t = T- \cdot\).)

In order to design energy minimization-type cost functionals below, we rewrite the differential part of the PIDE operator in divergence form:
\[\mathcal A u = \mathcal Lu + f[u]\]
with \(\mathcal L, f\) given by
\begin{align}
\mathcal Lu &=  -\sum_{j=1}^d\frac{\partial}{\partial x_j}\biggl( \sum_{i=1}^d a_{ij}(x)\frac{\partial u}{\partial x_i}\biggr)+ru, \\
f[u] &= \sum_{i=1}^d\biggl(b_i(x)+\sum_{j=1}^d\frac{\partial}{\partial x_j}a_{ij}(x)\biggr)\frac{\partial u}{\partial x_i} - I_\nu[u].
\end{align}
In the decomposition above, \(\mathcal L\) is a self-adjoint operator, while the remainder term $f[\cdot]$ comprises of both symmetric and non-symmetric components; in particular, the second part of the integral operator, \(\int_{\mathbb R^d} u(t,x)\nu(x,\dz)\), is symmetric. 
We chose to retain this in the remainder term for stability reasons of the proposed numerical framework below.

\subsection{Implicit-explicit minimizing movement method}\label{sec:ritz}

We will exploit now the divergence form of the PIDE in order to determine a minimizing movement approach, from which a respective cost functional will arise. 

We will first describe the method for the PIDE defined on $\mathbb{T}\times \Omega$ with $\mathbb{T}=(0,T]$ and \(\Omega =[0,x_{\mathrm{max}}]^d\), $x_\max>0$. 
In the next sections, we will also discuss specific modelling issues related to option pricing problems, which are, in turn, incorporated in order to extend the solution domain of the method. 
{ Let us, for the moment, prescribe homogeneous Dirichlet boundary conditions on $\partial\Omega$; this will be revisited in the next subsection.} 


We consider a subdivision of the time interval \((0,T]\) into \(n\) equally spaced time intervals $(t_{k-1},t_k]$ with \(t_k = k\tau,\ k=0,1,\dots,n\), for
\(\tau = T/n\). 
Let $u^0:=u_0(\cdot)$, then we seek approximations \(u^k\approx u(t_k,\cdot)\) by involving values of the \(p\) previous time steps.

In this context, we consider implicit-explicit backward differentiation formulae (BDF); we refer to \citet{Akrivis1999} for their numerical analysis. More specifically, for given parameter sets \( \beta_j, \gamma_j\), we consider the time-stepping methods
\begin{equation}
  \label{eq:implicit_explicit}
  \frac{\beta_p u^k-\sum_{j=0}^{p-1}\beta_j u^{k-j-1}}{\tau} +  \mathcal L u^k +  \sum_{j=0}^{p-1}\gamma_jf[u^{k-j-1}]=0,
\end{equation}
for $k=p,p+1,\dots,n$, along with special treatment of the first $p$ timesteps (\textit{e.g.}, by another, perhaps one-step, method).
In particular, for \(p=1\) we get the implicit-explicit Euler scheme
\begin{equation}
  \frac{u^k-u^{k-1}}{\tau} + \mathcal L u^k + f[u^{k-1}]=0,
\end{equation}
and for \(p=2\) the BDF-2 scheme
\begin{equation}
  \frac{\frac{3}{2}u^k-2 u^{k-1} + \frac{1}{2} u^{k-2}}{\tau} + \mathcal L u^k + 2f[u^{k-1}]-f[u^{k-2}]=0.
\end{equation}

A crucial observation, following from the minimizing movement point of view, is that \eqref{eq:implicit_explicit} (upon multiplication by $\tau$ and integration by parts) is the Euler--Lagrange equation of the convex minimization problem:
\begin{equation}\label{minimizing_movement}
\mathcal{C}[u]:=  \frac{1}{2}\Big\|\beta_p u-\sum_{j=0}^{p-1}\beta_ju^{k-j-1}\Big\|_{L^2(\Omega)}^2 + \tau \int_\Omega\mathcal{E}[u] \ud x
  + \tau \sum_{j=0}^{p-1} \gamma_j \int_\Omega f[u^{k-j-1}]u\, \d x\to \min,
\end{equation}
whereby
\begin{equation*}
    \mathcal{E}[u] := \frac{1}{2}\sum_{i,j =
    1}^d \Big(\alpha_{ij}(x)\frac{\partial u}{\partial x_i}\frac{\partial u}{\partial x_j} + ru^2\Big),
\end{equation*}
denotes the respective Dirichlet energy.

\subsection{Decomposition of the solution}

In the above discussion, we assumed, for simplicity, homogeneous Dirichlet boundary conditions on the domain of the truncated boundary $\partial\Omega$. 
In the option pricing context, however, one expects non-zero values of $u$ as $|x|\to\infty$. 
Fortunately, the modelling asserts that these values of the solution $u$ as $|x|\to\infty$ are asymptotically known. 
In particular, the value of the option at time \(t\geq 0\) can be expressed as
\begin{equation}\label{decomp}
  u(t,x) = w(t,x) + v(t,x),
\end{equation}
whereby \(w(t,x)\geq 0\) denotes the time value of the option, while \(v(t,x)\) denotes the intrinsic value of the option, given by
\begin{equation}
  u(t,x) \geq  \biggl(\sum_{i=1}^d \alpha_i x_i - \e^{-rt}\biggr)^+ =
  u_0(x) + (1-\e^{-rt})H\biggl(\sum_{i=1}^d \alpha_i x_i - \e^{-rt}\biggr)
  =: v(t,x),
\end{equation}
where \(H(\cdot)\) is the Heaviside function.

Crucially, for large values of the moneyness, \textit{i.e.} of \(\sum_{i=1}^d \alpha_ix_i\), the price of the option illustrates an asymptotic behavior
\begin{equation}\label{limit_inf}
  \lim_{\sum a_i x_i \rightarrow \infty}u(t,x) =\biggl(\sum_{i=1}^d \alpha_i x_i - \e^{-rt}\biggr)^+ =
  \sum_{i=1}^d \alpha_i x_i - \e^{-rt}.
\end{equation}
Therefore, the decomposition \eqref{decomp} implies that $\lim_{\sum a_i x_i \rightarrow \infty}w(t,x)=0$, which justifies the use of homogeneous Dirichlet boundary conditions on the domain $\Omega=[0,x_{\max}]^d$ for $x_{\max}$ large enough, if we solve for $w$ instead.

On a technical note, in order to avoid any issues due to the lack of smoothness of the Heaviside function, we mollify the lower-bound term by setting
\begin{equation}
  \label{eq:decomposition}
  \tilde v(t,x;\eta) = \biggl(\sum_{i=1}^d \alpha_i x_i - 1\biggr)^+
  + (1-\e^{-rt})\mathrm{Sigmoid}\Big(\sum_{i=1}^d \alpha_i x_i - \e^{-rt};\eta\Big),
\end{equation}
with
$  \mathrm{Sigmoid}(x;\eta) := \big(1+\e^{-\eta x}\big)^{-1}$, $\eta >0$.
Obviously,
\(\lim_{\eta\rightarrow\infty}\tilde v(t,x;\eta) =  v(t,x)\).
Moreover, the parameter \(\eta\) influences the smoothness of the gradient of \(\tilde v(t,x;\eta)\) with respect to \(x\). 
In particular, smaller values of \(\eta\) result in smoother approximations.
Hence, the parameter \(\eta\) should be chosen appropriately, to strike a balance between the precision of the lower bound approximation and the desired level of smoothness.

\subsection{Domain truncation}

Motivated by the discussion above, we now provide a framework for the truncation of the spatial domain and the extension to values in $\mathbb{R}^d_+$ that reflects the natural setting of the problem. 
To that end, recall \eqref{limit_inf} and observe that 
\[
    \biggl(\sum_{i=1}^d \alpha_i x_i - \e^{-rt}\biggr)^+\quad  \text{ is parallel to }\quad \sum_{i=1}^d \alpha_i x_i.
\]
Consequently, for \(x^{\prime}\)  associated with a higher moneyness level than $x$, we may establish a connection between the option prices inside the truncated domain and those of ``far away'' values using the following expression:
\begin{equation}
  \label{eq:approx_large_moneyness}
  u(t,x^{\prime}) \approx u(t,x) + \sum_{i=1}^d \alpha_i (x_i^{\prime}-x_i),
\end{equation}
for \(\sum_{i=1}^d a_i x_i\) sufficiently large.

Let \(\mathcal{R}\) denote a subset of \(\mathbb{R}^d_+\) such that \(\sum_{i=1}^d \alpha_i x_i < x_r\) and \(x_i \leq x_{\mathrm{max}},\ i\in\mathbb I\), for a given \(x_r \leq x_{\mathrm{max}}\).
The boundary \(\partial \mathcal R\) of \(\mathcal R\) consists of the points which satisfy \(\sum_i \alpha_i x_i = x_r\) and \(x_i \leq x_{\mathrm{max}}\) for \(i\in\mathbb I\).

\begin{figure}
    \centering
    \includegraphics[width=0.5\linewidth]{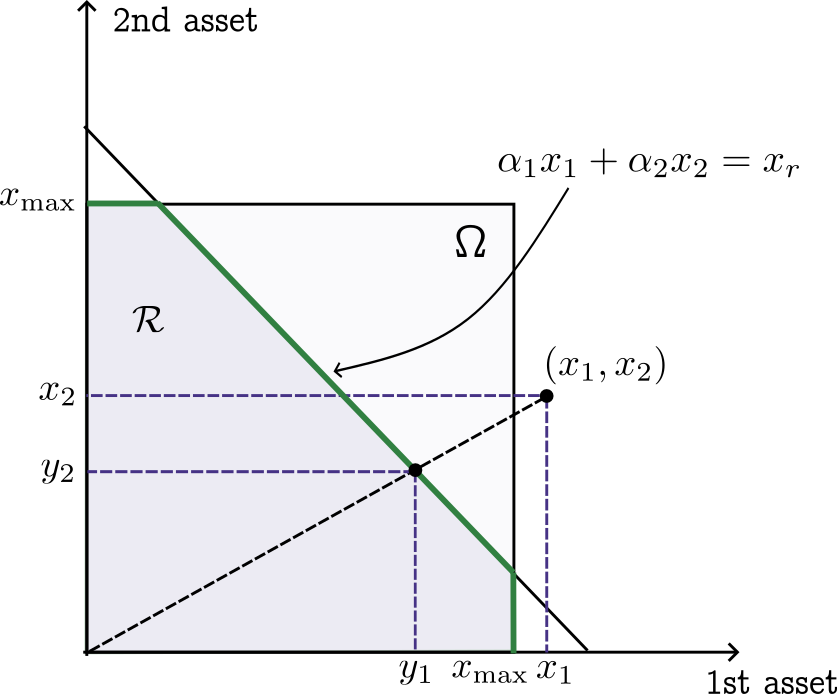}
    \caption{A two-dimensional example for the domain truncation. The green boundary defines the projection surface for the option price in extreme moneynesses.}
    \label{fig:ext2d}
\end{figure}

We define the projection \(y=(y_1,y_2,\dots,y_d)\) of \(x \in \mathbb{R}^d_+\) as
\(y_i = q(x)x_i\), with
\begin{equation}\label{projection}
q(x) =
    \begin{cases}
        x_{\mathrm{max}}/\mathrm{max}\{x_i\},& \text{if } \ \mathrm{max}\{x_i\} \geq \mathrm{max}\biggl(\sum_{i=1}^d \alpha_i x_i,x_r\biggr)x_{\mathrm{max}}/x_r\\
        x_r/\mathrm{max}\biggl(\sum_{i=1}^d \alpha_i x_i,x_r\biggr),& \text{otherwise.}
    \end{cases}
\end{equation}
Clearly, \(0 \leq q(x) \leq 1\) for all \(x\in \mathbb R^d_+\) and  \(q(x) x \in \mathcal R\). 
Employing the approximation \eqref{eq:approx_large_moneyness}, we
introduce a formula that expresses the option prices for each \(x\in \mathbb R^d_+\) in relation to the option price within the bounded domain \(\mathcal R\):
\begin{equation}
  u(t,x) \approx u(t,y(x)) + \sum_{i=1}^d \alpha_i(x-y(x)), \quad \text{for each } \ x \in \mathbb R^d_+.
\end{equation}
Figure \ref{fig:ext2d} illustrates the proposed domain truncation in a two-asset scenario. In this example, \(x\) has a higher moneyness than \(x_r\), resulting in \(y = (y_1,y_2)\) of moneyness  \(x_r\), by following the projection procedure.

The significance of the domain truncation lies in getting estimates of the option price for large values of the moneyness, even within the confines of a solution in a bounded domain. This is crucial considering the random discontinuities in asset prices that we have taken into account.

%% file: modeling.tex
In this section, we detail the representation of the solution through a deep artificial neural network (ANN) and discuss the network parameter optimization process. 
A key attribute of the methodology presented below is that the approximate solution $U^k$ is computed at each timestep $t_k$ by a deep ANN with a specific architecture; this has been proposed, for instance, by \citet{Georgoulis2023}, see also the deep BSDE method of \citet{Han_Jentzen_E_2018}, and is in contrast to other popular ANN-type methods for PDEs, such as the deep Galerkin method (DGM) of \citet{Sirignano2018} or the physics-informed neural networks (PINNs) by \citet{pinn_giannena,PINN}, whereby a single ANN approximating the solution over a space-time cylinder is employed. 
A practical advantage of using a single deep ANN architecture for each time instance $t_k$ is that the approximation is computed by retraining the same architecture from one timestep to the next, thereby reducing the training time due to the availability of good ``starting'' parameter values, \textit{i.e.}, the values from the previous time-step.

\subsection{Network architecture}\label{net_arch}

\begin{figure}
\begin{center}
    \includegraphics[width=0.85\linewidth]{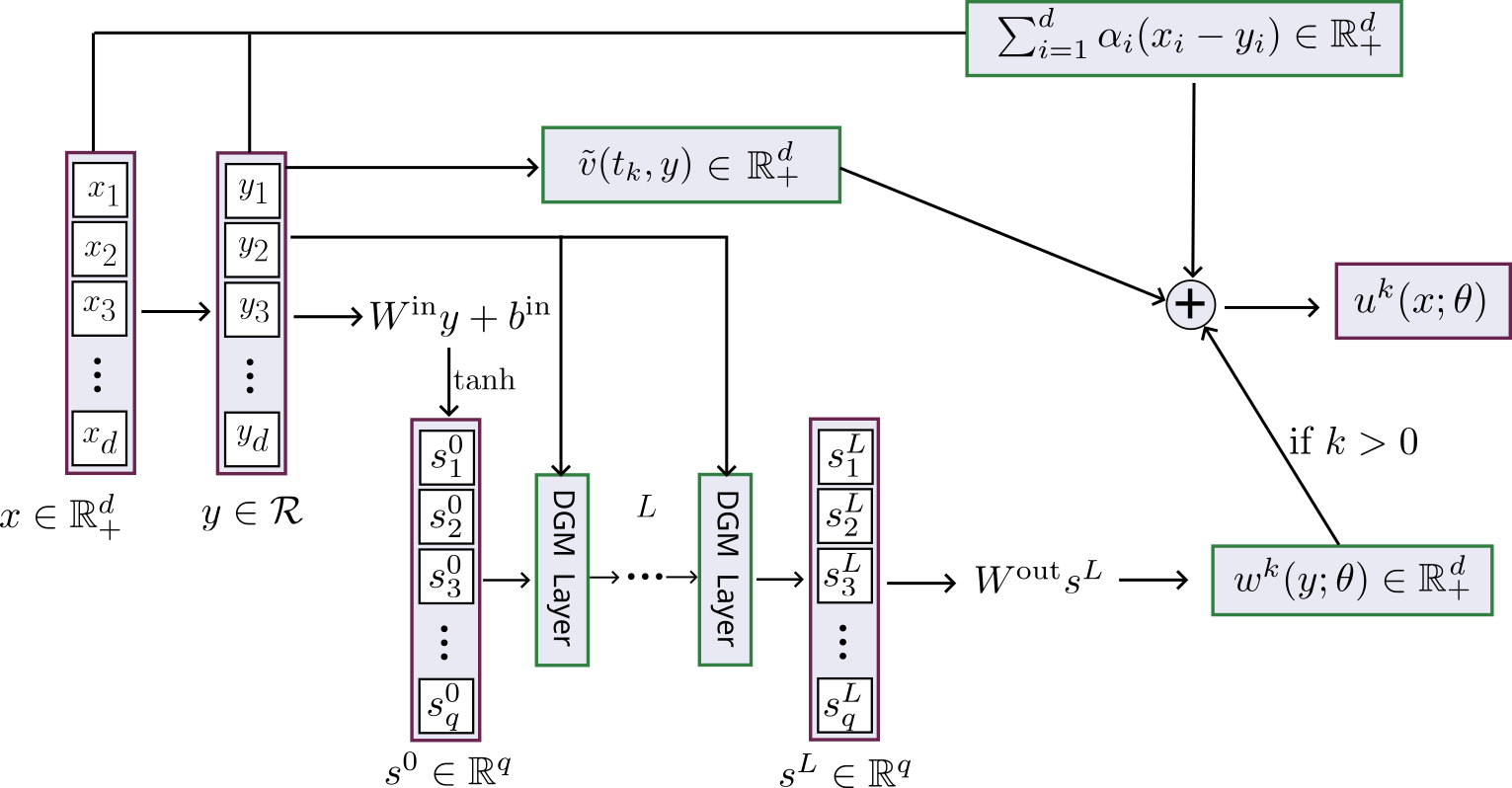}
    \caption{Flowchart of the deep neural network modeling the solution of the PIDE for a time instance \(t=t_k\).}
    \label{fig:NN}
\end{center}
\end{figure}

\begin{figure}
\begin{center}
    \includegraphics[width=0.75\linewidth]{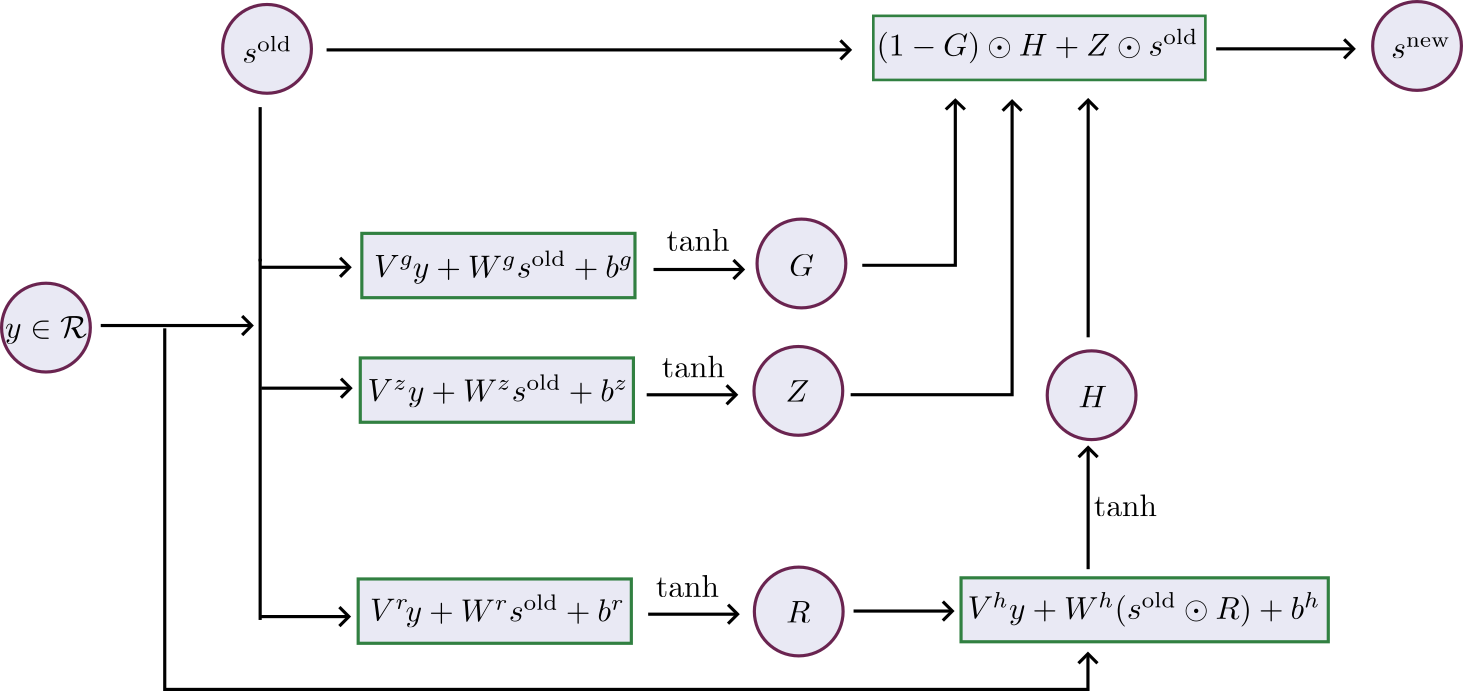}
    \caption{The architecture of a DGM layer.}\label{fig:dgm_layer}
\end{center}
\end{figure}

The approximate solution $U^k$, for each timestep $t_k$, is represented by a modified version of the deep ANN of residual type by \citet{Sirignano2018}; the architecture of each layer follows \cite{Sirignano2018} and, for this reason, is henceforth designated as ``DGM layer''. 
The network implements the decomposition and boundary modeling described in the previous subsections. 
More specifically, for an input $y$, we set
\begin{align*} 
S^0 &= \tanh(W^{\mathrm{in}} y + b^{\mathrm{in}}),  \\
\text{ DGM layer}\\
|\quad G^\ell &= \tanh(V^{g,\ell} y + W^{g,\ell} S^{\ell-1} + b^{g,\ell}),\ \ell = 1,\dots, L \\
|\quad 
Z^\ell &= \tanh(V^{z,\ell} y + W^{z,\ell} S^{\ell-1} + b^{z,\ell}),\ \ell = 1,\dots, L \\
|\quad R^\ell &= \tanh(V^{r,\ell} y + W^{r,\ell} S^{\ell-1} + b^{r,\ell}),\ \ell = 1,\dots, L \\
|\quad\! H^\ell &= \tanh(V^{h,\ell} y + W^{h,\ell}( S^{\ell-1} \odot R^{\ell}) + b^{h,\ell}),\ \ell = 1,\dots, L \\
\lfloor\quad S^{\ell} &= (1-G^{\ell})\odot H^{\ell} + Z^{\ell}\odot S^{\ell-1},\ \ell = 1,\dots, L\\
\tilde v(t_k,y) &= \biggl(\sum_{i=1}^d \alpha_i y_i - 1\biggr)^+ + (1-\e^{-rt_k})\ \mathrm{Sigmoid}\Big(\sum_{i=1}^d \alpha_i y_i-\e^{-rt_k};\eta\Big),\\
w^k(y;\theta) &= \mathrm{Softplus}\big(W^{\mathrm{out}}S^L;\delta\big),
\end{align*}
with \(L\) denoting the number of hidden layers and \(\odot\) denoting the Hadamard product. 
The trainable parameters \(\theta\) of the model are
\begin{equation}
  \theta = \{W^{\mathrm{in}},b^{\mathrm{in}}, (V^{*,\ell},W^{*,\ell},b^{*,\ell})_{\ell=1,\dots,L}^{*\in\{g,z,r,h\}},W^{\mathrm{out}}\},
\end{equation}
and the output of the network is given by
\begin{equation}
 U(t_k,x;\theta) = w^k(y;\theta) + \tilde v(t_k,y) + \sum_{i=1}^d \alpha_i (x_i - y_i),
\end{equation}
recalling the definition of $y\equiv y(x)$ (and of the $y_i$'s) in \eqref{projection}.
Figure \ref{fig:NN} shows a flowchart of the adopted `global' ANN architecture, while Figure \ref{fig:dgm_layer} presents the architecture of a single DGM layer. 

The computation of the parameters $\theta$ which provide the approximate solution at $t_k$ is performed using \eqref{minimizing_movement}, \textit{i.e.}, we seek $\theta$ minimizing the cost functional
\begin{equation}\label{continuous_cost}
\begin{aligned}
\mathscr{C}_k(\theta) 
    := \mathcal{C}[U(t_k,\cdot,\theta)]
&= \ \frac{1}{2} \Big\| \beta_p U(t_k,\cdot,\theta)-\sum_{j=0}^{p-1}\beta_jU(t_{j(k)},\cdot,\theta^{j(k)}) \Big\|_{L^2(\Omega)}^2 \\
&\quad + \tau \int_\Omega \mathcal{E}[U(t_k,x,\theta)] \, \d x
  + \tau \sum_{j=0}^{p-1} \gamma_j \int_\Omega f[U(t_{j(k)},x,\theta^{j(k)})]U(t_k,x,\theta)\, \d x,
  \end{aligned}
\end{equation}
using the notation $j(k):=k-j-1$ for brevity.
Let us point out that, due to the network architecture, although $\mathcal{C}$ is convex with respect to its argument, $\mathscr{C}$ is not, in general. 
The minimizer is denoted by $\theta^k$. 
Due to the multistep nature of the BDF methods, we observe that $\theta^k$ also depends on $\theta^{k-1},\dots, \theta^{k-p}$. 
As discussed above, special care has to be taken for the first $p$ time steps (\textit{e.g.}, by employing an one-step method) to initiate the iteration. 

\begin{remark}
We note the practical significance of mollifying the solution lower bound as per \eqref{eq:decomposition}. Indeed, 
$v(t_{j(k)},\cdot) $ is contained in $U(t_{j(k)},\cdot,\theta^{j(k)})$ and it is differentiated in space within $\mathscr{C}$.
\end{remark}

\subsection{Training}

As is typical in ANN-based methods for PDEs, the cost function \eqref{continuous_cost} is discretized by a Monte Carlo sampling procedure. 
In particular, considering  \(N\) uniformly sampled points \(\{x^i\}_{i=1}^N\) in \(\Omega = [0,x_{\mathrm{max}}]^d\), the discretized cost functional is given by
\begin{multline}
\label{eq:cost_functional}
\widetilde{\mathscr{C}}_k(\theta)
    :=\frac{(x_{\mathrm{max}})^d}{N} \sum_{i=1}^N \biggl\{ \frac{1}{2}\Big[\beta_p U(t_k,x^i,\theta) 
        - \sum_{j=0}^{p-1}\beta_jU(t_{j(k)},x^i,\theta^{j(k)})\Big]^2  \\
    + \tau \mathcal{E}[U(t_k,x^i,\theta)] + \tau \sum_{j=0}^{p-1} \gamma_j f[U(t_{j(k)},x^i,\theta^{j(k)})]U(t_k,x^i,\theta) \biggl\}, 
\end{multline}
using again $j(k):=k-j-1$ for brevity.
For notational simplicity, we still denote by $\theta^k$ the computed minimizer of $\widetilde{\mathscr{C}}_k(\theta)$. In practice, we update the model parameters
using the ADAM optimizer for \(N_k\) training epochs and we denote by $\theta^k$ the result of the optimization (even if it may not have converged).

\subsubsection{Initialization}

The initial condition \eqref{eq:PIDE} implies that \(w^0(x;\theta^{0}) = 0\) for all \(x\in [0,x_r]^d\), which may result in practical difficulties in the training process, since it leads to vanishing weights.
In order to address this, we use the mathematically equivalent form
\begin{equation*}
 U(t_k,x;\theta) = (1-\delta_{k0})w^k(y;\theta) + \tilde v(t_k,y) +
 \sum_{i=1}^d \alpha_i (x_i - y_i),
\end{equation*}
with \(\delta_{ij}\) being the Kronecker delta. In other words, we impose \(w^0(x;\theta^{0}) = 0\) strongly at $t=0$, resulting in the network weights becoming independent of the initial condition of the PIDE.

Nevertheless, it is of interest to initialize \(w^0\) in order to get a good starting point for adapting the network for approximating the solution at \(t_1\) and beyond. 
In that respect, we employ the following smooth function that goes to zero for small and large asset prices:
\begin{equation*}
  f^0(x) = \epsilon
  \exp\biggl[-\frac{1}{2\zeta^2(x)}\biggl(\sum_{i=1}^d\alpha_ix_i-1\biggr)^2 \
  \biggr],
\end{equation*} 
for $\epsilon>0$ small, with \(\zeta(x)\) taking two possible positive values \(\zeta_1,\zeta_2\) according to the criterion \(\sum_{i=1}^d \alpha_i x_i > 1\)  and we compute $\theta^0$ such that
\begin{equation*}
  w^0(x;\theta^0)\to \min_\theta \big\|w^0(x;\theta)-f^0(x)\big\|^2_{L^2([0,x_r]^d)}.
\end{equation*}
Once computed, we employ $ w^0(x;\theta^0)$ in the BDF iteration.

%% file: integral.tex
Each evaluation of the cost (loss) function requires the numerical calculation of the integral term $I_\nu[u]$ in $f[u]$.
Approximating the integral operator $I_\nu$ requires special treatment due to the complexity implied by this operator. 
To that end, we calculate the integral operator using two methods. 
The first one employs a sparse-grid integration technique based on the Gauss--Hermite quadrature, while the second one involves training a specialized neural network for computing the integral values. The latter method is more time-efficient, whereas the former gives, in general, more accurate results.

The integral term appears in an explicit fashion with respect to the timestepping, therefore we are required to evaluate the values $I_\nu[U(t_{j(k)},x,\theta^{j(k)})]$ at $x$ for known ANN solution approximations $U(t_{j(k)},x,\theta^{j(k)})$, $j=0,\dots,p-1$, again for $j(k)=k-j-1$.

\subsubsection{Sparse Gauss--Hermite quadrature}

In the Merton model, the integral can be expressed as the expected value, with respect to a Gaussian measure, of the function \(h(z)=u(x\e^z)-u(x)\) multiplied by the Poisson parameter \(\lambda\); for brevity, for the remainder of this discussion we suppress the dependence on $t$, and $u$ signifies any of the $U(t_{j(k)},x,\theta^{j(k)})$ for $j=0,\dots,p-1$. 
In addition, the amplitude of the random jumps is determined by a multivariate normal random variable, \(Z\sim \mathcal N (\mu_J,\Sigma_J)\). 
Therefore, $\Sigma_J$ introduces directional features. 

\begin{figure}[h]
  \centering
  \includegraphics[scale=0.75]{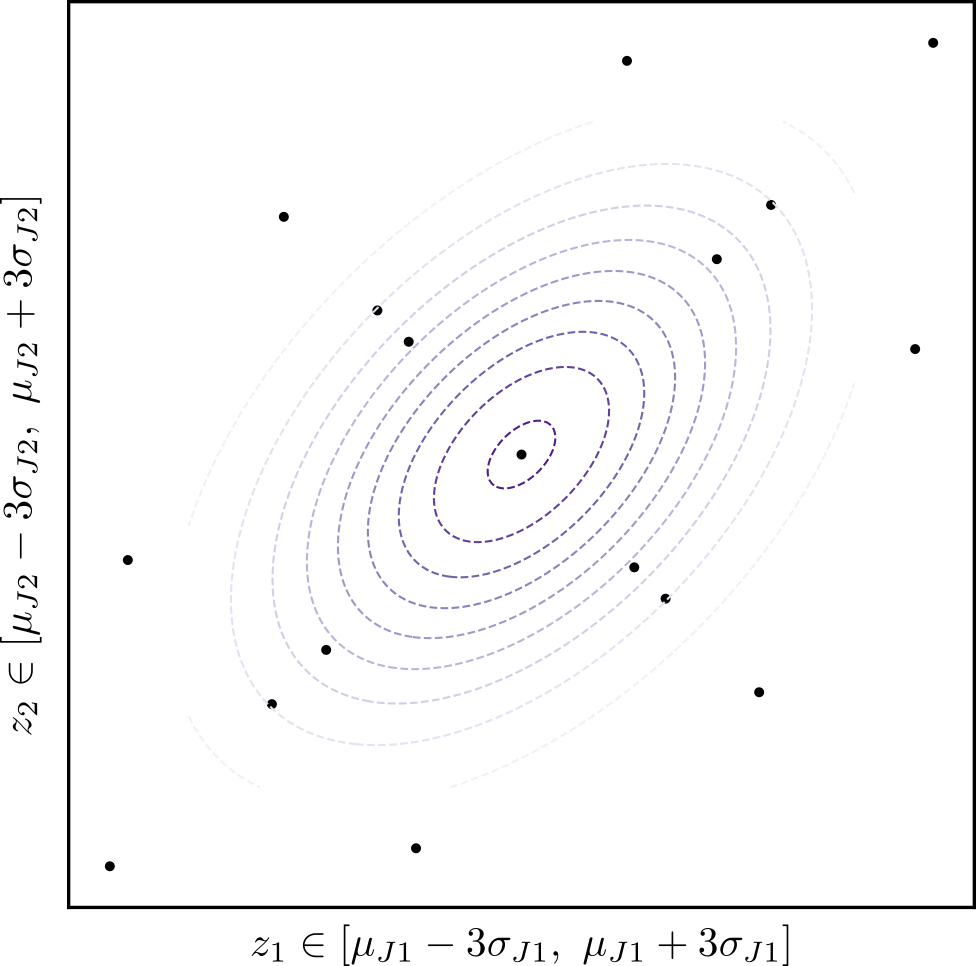}
  \caption{\label{fig:sparse2d} Sparse sampling in \(\mathbb R^2\) using the Gauss--Hermite quadrature for normally distributed random variables with correlation equal to $\frac12$.}
\end{figure}

The construction of a $d$-dimensional quadrature rule employing sparse grid interpolation of the integrand, benefits by following the ``natural direction'' of $\Sigma_J$. 
To that end, we first perform the Singular Value Decomposition (SVD) of the covariance matrix to express the integration in terms of the Gauss--Hermite quadrature by modulating and rotating the original axes, viz.,
\begin{equation}
  \Sigma_J = A\Lambda A^{\mathsf{T}} = A \Lambda^{1/2}\Lambda^{1/2}A^{\mathsf{T}}=BB^{\mathsf{T}},
\end{equation}
where \(B=A\Lambda^{1/2}\). 
The change of variables \(Z - \mu = \sqrt{2}B Y\) transforms the integrand such that the multidimensional Gauss--Hermite quadrature can produce a sparse grid of representative points. 
More specifically, we have
\begin{align}
\int_{\mathbb{R}^d}h(z)\nu(\dz) 
    &= \lambda\pi^{-d/2} \int_{\mathbb{R}^d} h( \mu + \sqrt{2}By) \exp(-y^{\mathsf{T}}y) \dy \nonumber \\
    &\approx \lambda\pi^{-d/2} \sum_{\mathbf i \in  \Theta_q}h(\mu + \sqrt{2}By^{\mathbf i})W^{\mathbf i},
\end{align}
with \(y^{\mathbf i}=(y^{i_1},\dots,y^{i_d})^{\mathsf{T}}\) representing a node of the sparse grid, and \(W^{\mathbf{i}} = \prod_{k=1}^d w^{i_{k}}\) the associated weight for a multi-index ${\mathbf i}=(i_1,\dots,i_d$). 
Here, \(y^{i_k}\) are chosen to be the roots of Hermite polynomials, and the index space \(\Theta_q\subset \mathbb{N}^d\) is selected in such a way that the resulting quadrature exactly integrates polynomials up to a desired degree $q$; \textit{cf.} \citet{griebel}. 
Figure \ref{fig:sparse2d} presents an example of a sparse grid of points in \(\mathbb R^2\) for integrating a function of two correlated normally distributed random variables.

\subsubsection{Approximation of the integral using an ANN}

We further develop an alternative approach for the evaluation of the integral operator, by introducing a second artificial neural network for this purpose. 
The introduction of a second ANN is inspired by \citet{Gnoatto2022}, however, we introduce a neural network that directly learns the values of the integral $\int h(z)\nu(\dz)$, while in \cite{Gnoatto2022} they learn the values of a martingale stemming from stochastically integrating $h$ with respect to the compensated Poisson measure of jumps. 

The neural network we now consider will serve as an estimator of the values of the integral $\int h(z)\nu(\dz)$ with $h(z)=u(x\e^z)-u(x)$ for each value of \(x\). 
The approximate neural network value at time \(t_k\) is denoted by \(\mathcal I^k(x;\phi)\), where \(\phi\) is the set of trainable parameters. 
In order to determine \(\phi^k\), we solve the optimization problem
\begin{equation}\label{opt_integral}
  \min_{\phi\in \Phi} \mathbb E \biggl[\mathcal I^k(x;\phi) - \sum_{j=1}^{p-1}\gamma_jI_\nu[U(t_{j(k)},x;\theta^{j(k)})]\biggr]^2,
\end{equation}
with the expectation taken over \(x\). 

As usual, we discretize the integral in \eqref{opt_integral} by Monte Carlo sampling before performing the optimization, \textit{i.e.}, we approximate an unbiased and minimum variance estimator of the integral operator with respect to \(x\). 
To that end, setting $h_{j(k)}(x,z):= U(t_{j(k)},x\e^{ z},\theta^{j(k)})-U(t_{j(k)},x,\theta^{j(k)})$, we consider normally distributed points $\{z^r\}_{r=1}^M$ in $\mathbb{R}^d$ according to the distribution that governs the size of the jumps, and seek to find the optimizer $\phi^k\in\Phi$ that minimizes
\begin{equation}
\label{eq:integral_min}
  \min_{\phi \in \Phi} \mathbb E \biggl[\mathcal I^k(x;\phi) -  \frac{\lambda}{M}\sum_{r=1}^M\sum_{j=1}^{p-1}\gamma_jh_{j(k)}(x,z^r)\biggr]^2;
\end{equation}
again, the expectation is over \(x\). 
At the time step \(t^k\) and given a set of uniformly distributed points \(\{x^i\}_{i=1}^N\), we obtain a discrete approximation of \eqref{eq:integral_min}:
\begin{equation}
\label{eq:cost_function_integral_term}
  \frac{(x_{\mathrm{max}})^d}{N}\sum_{i=1}^N \biggl[\mathcal I^k( x^i;\phi) - \frac{\lambda}{M}\sum_{r=1}^M \sum_{j=1}^{p-1} \gamma_j h_{j(k)}(x^i,z^r)\biggr]^2. 
\end{equation}

Concluding, in order to estimate the solution $U^k$ of the PIDE at time \(t^k\), we need to determine the values of the integral operator. To approximate these values, we employ the supplementary ANN. 

At each time step, we begin by optimizing the parameters \(\phi^k\) of the ANN for the integral by minimizing \eqref{eq:cost_function_integral_term}, followed by the independent optimization for \(\theta^k\) based on \eqref{eq:cost_functional}.
Fortunately, the training associated with \eqref{eq:cost_function_integral_term} is computationally inexpensive, introducing no significant overhead to the overall scheme.



%% file: applications.tex
We shall now test the performance of the proposed methodology by pricing a European basket call option in the Merton model.  
Also, we provide a simple comparison with two popular and successful machine learning-driven solvers.
More specifically, we consider a basket of \(d\) equally weighted assets with moneyness values \(x_1, x_2,\dots,x_d\).
The model parameters are chosen to be the following:
\[
    \sigma_i=0.5,\ \rho_{ij} = \delta_{ij} + 0.5(1-\delta_{ij}),\ \ i,j\in\mathbb{I},\ t \in \mathbb T=(0,T],
\]
where \(\delta_{ij}\) denotes the Kronecker delta again. 
Moreover, the parameters of the jump distribution are
\[
    \lambda =  1, \ \mu_{J_i}=0,\ \sigma_{J_i}=0.5,\ \rho_{J_{ij}} = \delta_{ij} + 0.2(1-\delta_{ij}),\ i,j\in\mathbb{I}.
\]
Note that the elements of the covariance matrices are given in terms of the
standard deviations and correlations as follows:
\[
    \Sigma_{ij}=\sigma_i\sigma_j\rho_{ij} \quad \text{ and } \quad \Sigma_{J_{ij}}=\sigma_{J_i}\sigma_{J_j}\rho_{J_{ij}}.
\]
Let us point out that we have intentionally selected large values for the volatilities and asset correlations, in order to test the performance of the method in challenging scenarios.
Below, we present numerical experiments for dimensions \(d=5\) and $d=15$.

\subsection{5 correlated assets -- Integration using Gauss--Hermite quadrature}

\begin{figure}
\begin{center}
    \includegraphics[width=\linewidth]{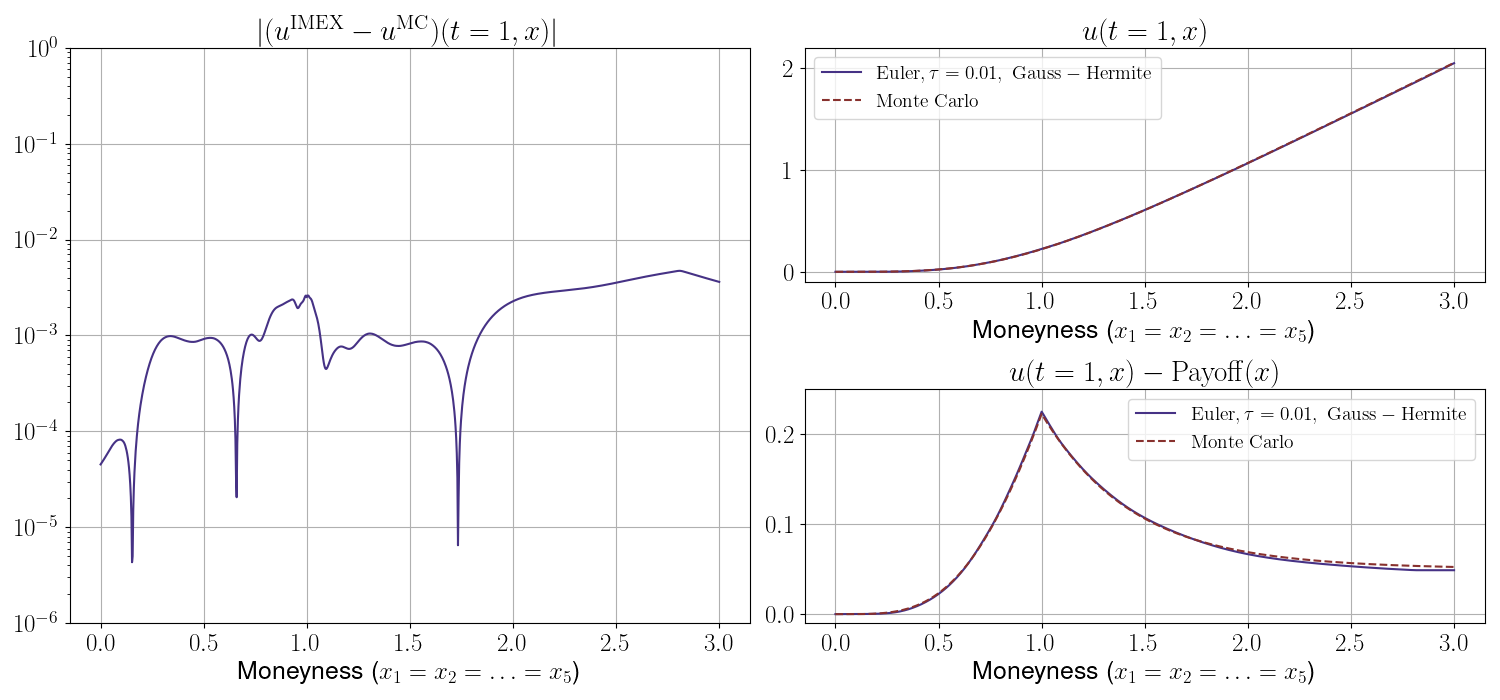}
    \caption{Basket option prices (top right), differences between price and payoff (bottom right) and differences between the proposed method and quasi Monte Carlo (left), using the implicit-explicit Euler scheme and the Gauss--Hermite quadrature for the integral, for \(d=5\) . }\label{fig:5_euler_gh.png}
\end{center}
\end{figure}

\begin{figure}
\begin{center}
    \includegraphics[width=\linewidth]{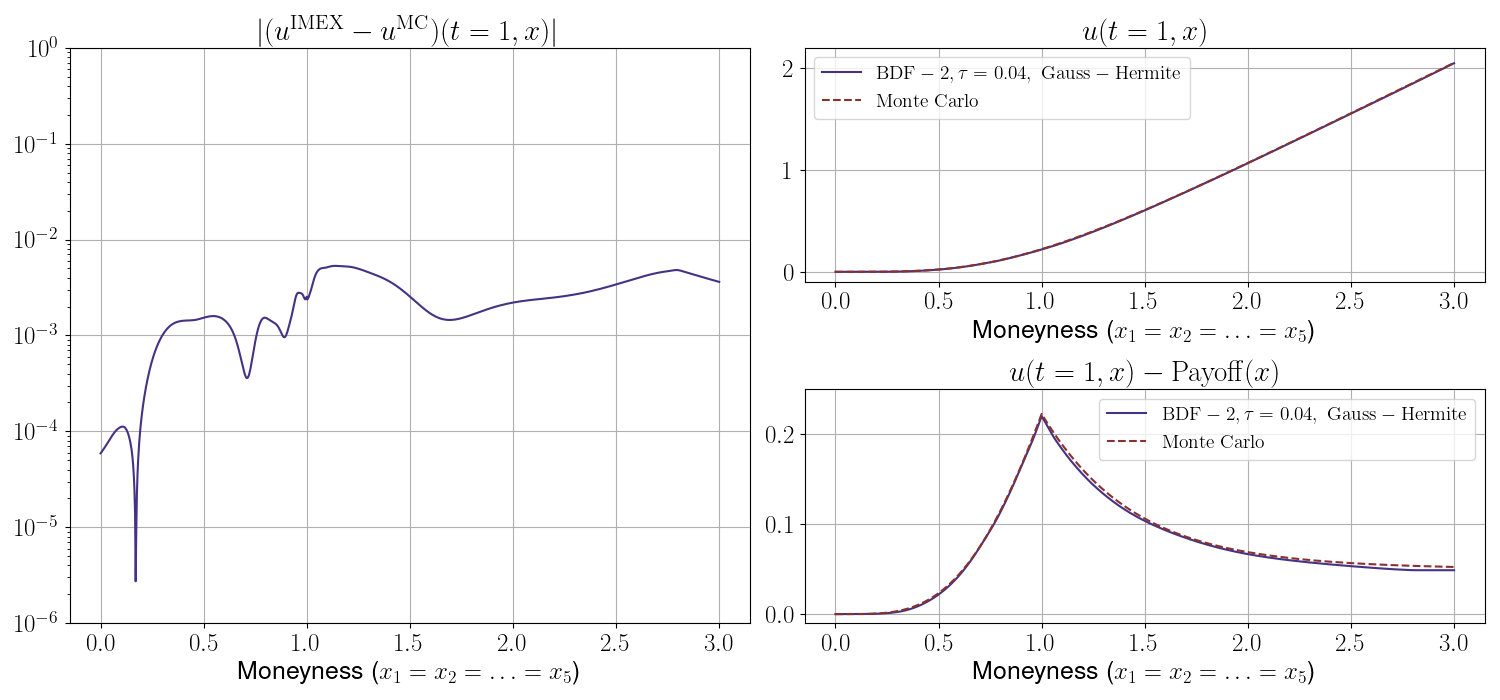}
    \caption{Basket option prices (top right), differences between price and payoff (bottom right) and differences between the proposed method and quasi Monte Carlo (left), using the BDF-2 scheme and the Gauss--Hermite quadrature for the integral, for $d=5$.}\label{fig:5_bdf2_gh.png}
\end{center}
\end{figure}

The first numerical example concerns the valuation of a European basket call option consisting of \(5\) correlated assets. 
The maturity is chosen to be in one year, \textit{i.e.} \(T=1\).
Initially, we use the Gauss--Hermite quadrature to compute the integral operator. 
Results for both the implicit-explicit Euler and the BDF-2 schemes are provided. 
For the Euler scheme, we consider a time-step of \(\tau = 0.01\), while the BDF-2 scheme uses a larger time-step of \(\tau=0.04\). 
We present the option prices for moneyness values in the interval \([0,3]\), considering the scenario where all assets share the same moneyness.
Figure \ref{fig:5_euler_gh.png} corresponds to the implicit-explicit Euler scheme, and Figure \ref{fig:5_bdf2_gh.png} to the BDF-2 scheme.
For comparison reasons, we also provide an extended Quasi Monte Carlo (QMC) estimation of the solution.
We can observe that both methods work very well compared to the QMC simulation, and the difference to the Quasi Monte Carlo is only visible for deep in-the-money options, with moneyness level above $2$.
Indeed, the absolute error is on the order of $10^{-3}$.
The implicit-explicit Euler method appears slightly more accurate than the BDF-2 scheme, due to the smallest time-stepping. 
This comes at a computational cost, since it is approximately $1.8$ times slower.

\subsection{5 correlated assets -- Integration using an ANN}

\begin{figure}
\begin{center}
    \includegraphics[width=\linewidth]{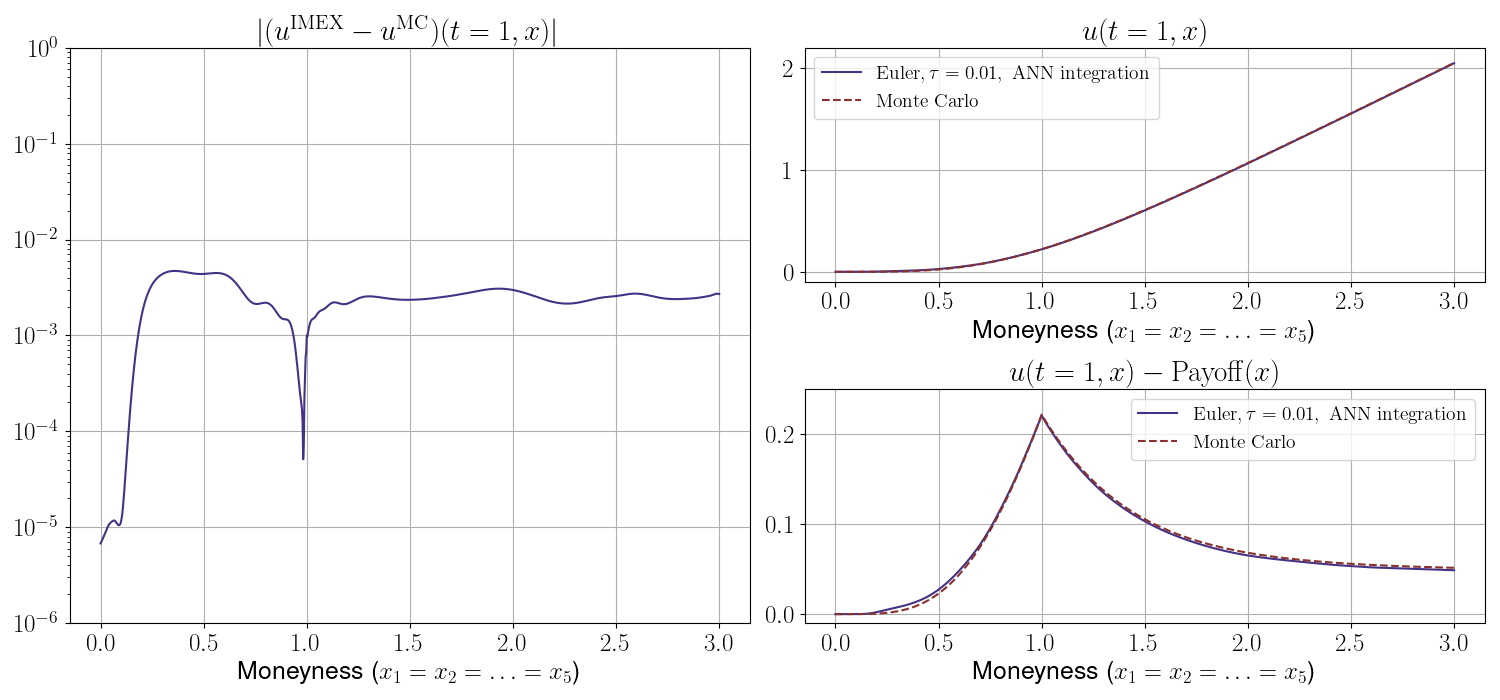}
    \caption{Basket option prices (top right), differences between price and payoff (bottom right) and differences between the proposed method and quasi Monte Carlo (left), using the implicit-explicit Euler scheme and an ANN for the calculation of the integral, for $d=5$.}\label{fig:5_euler_auto.png}
\end{center}
\end{figure}

\begin{figure}
\begin{center}
    \includegraphics[width=\linewidth]{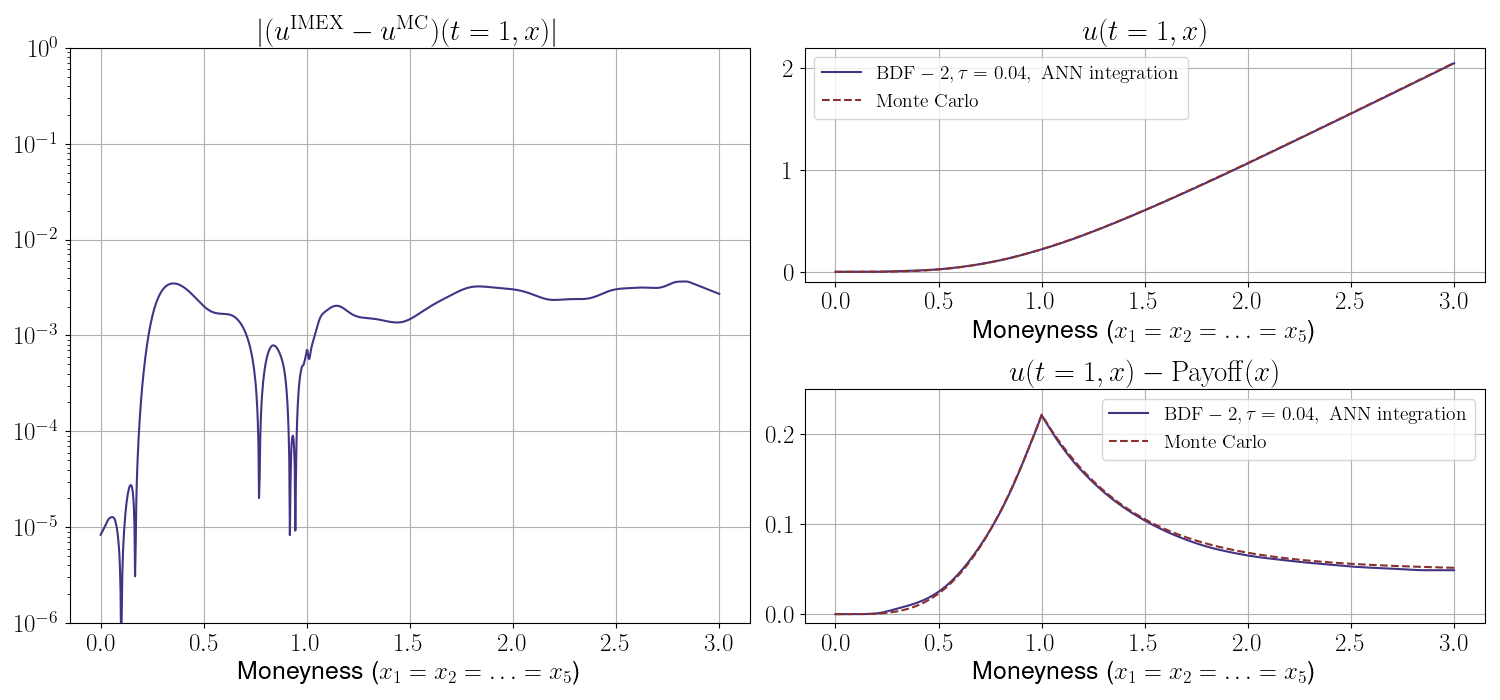}
    \caption{Basket option prices (top right), differences between price and payoff (bottom right) and differences between the proposed method and quasi Monte Carlo (left), using the BDF-2 scheme and an ANN for the calculation of the integral operator, for $d=5$.}\label{fig:5_bdf2_auto.png}
\end{center}
\end{figure}

Next, we present the corresponding results using the second ANN instead of the Gauss--Hermite quadrature to estimate the integral operator.
Figures \ref{fig:5_euler_auto.png} and \ref{fig:5_bdf2_auto.png} depict the recovered option prices for the implicit-explicit Euler and the BDF-2 schemes, respectively.
Again, we compare the results using a greedy Quasi Monte Carlo simulation.
We can observe that the ANN computation of the integral operator also works very well, and the absolute error remains in the order of $10^{-3}$.
The ANN method though results in a computation of basket options prices that is roughly $1.7$ times faster than the corresponding method with the Gauss--Hermite quadrature.

\begin{figure}
\begin{center}
    \includegraphics[width=0.8\linewidth]{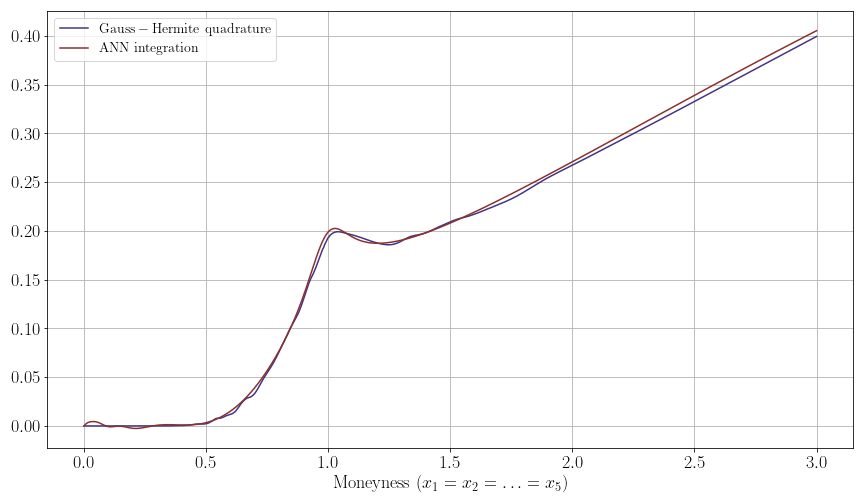}
    \caption{Estimates of the integral operator for \(t=0.01\).}
    \label{fig:integral01.png}
\end{center}
\end{figure}

Figure \ref{fig:integral01.png} illustrates a comparison of the two approaches employed for the computation of the integral operator in the case of the 5-asset European basket option for moneynesses in \([0,3]\) at \(t=0.01\). 
We can see that the two methods yield similar approximations of the integral operator. 
Let us point out that the Gauss--Hermite quadrature tends to outperform the ANN for extreme moneyness values (small or large) due to its sampling-independent nature. 
More specifically, the quadrature captures the asymptotic behavior of the integral for large \(x\), aligning with the expected jump sizes, \textit{i.e.}: \(\lim_{x\rightarrow \infty} I_{\varphi}(x) = x\mathbb E [\e^z-1]\). 
On the other hand, the ANN-based approach tends to yield smoother approximations, particularly for commonly sampled moneyness levels.

\subsection{15 correlated assets -- Integration using an ANN}

\begin{figure}
\begin{center}
    \includegraphics[width=\linewidth]{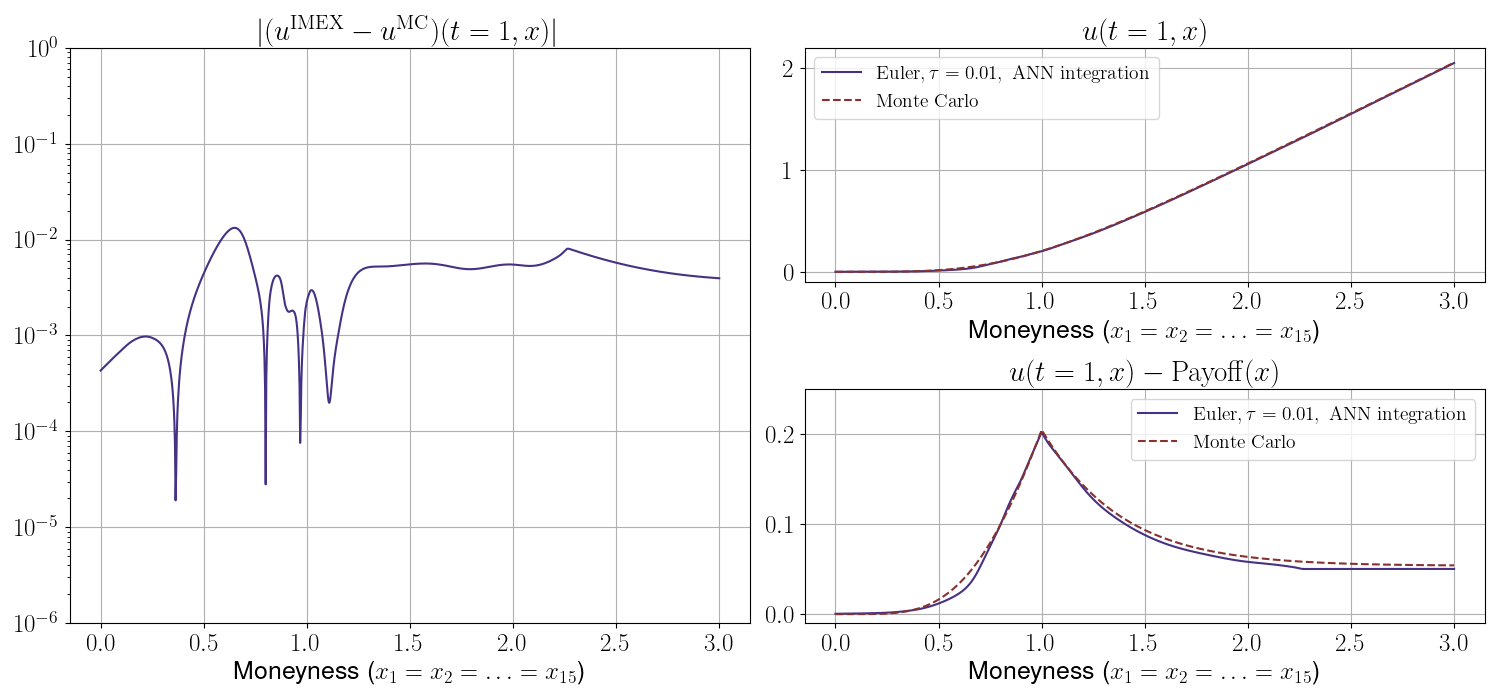}
    \caption{Basket option prices (top right), differences between price and payoff (bottom right) and differences between the proposed method and quasi Monte Carlo (left), using the implicit-explicit Euler scheme and an ANN for the calculation of the integral, for $d=15$.}\label{fig:15_euler_auto.png}
\end{center}
\end{figure}

\begin{figure}
\begin{center}
    \includegraphics[width=\linewidth]{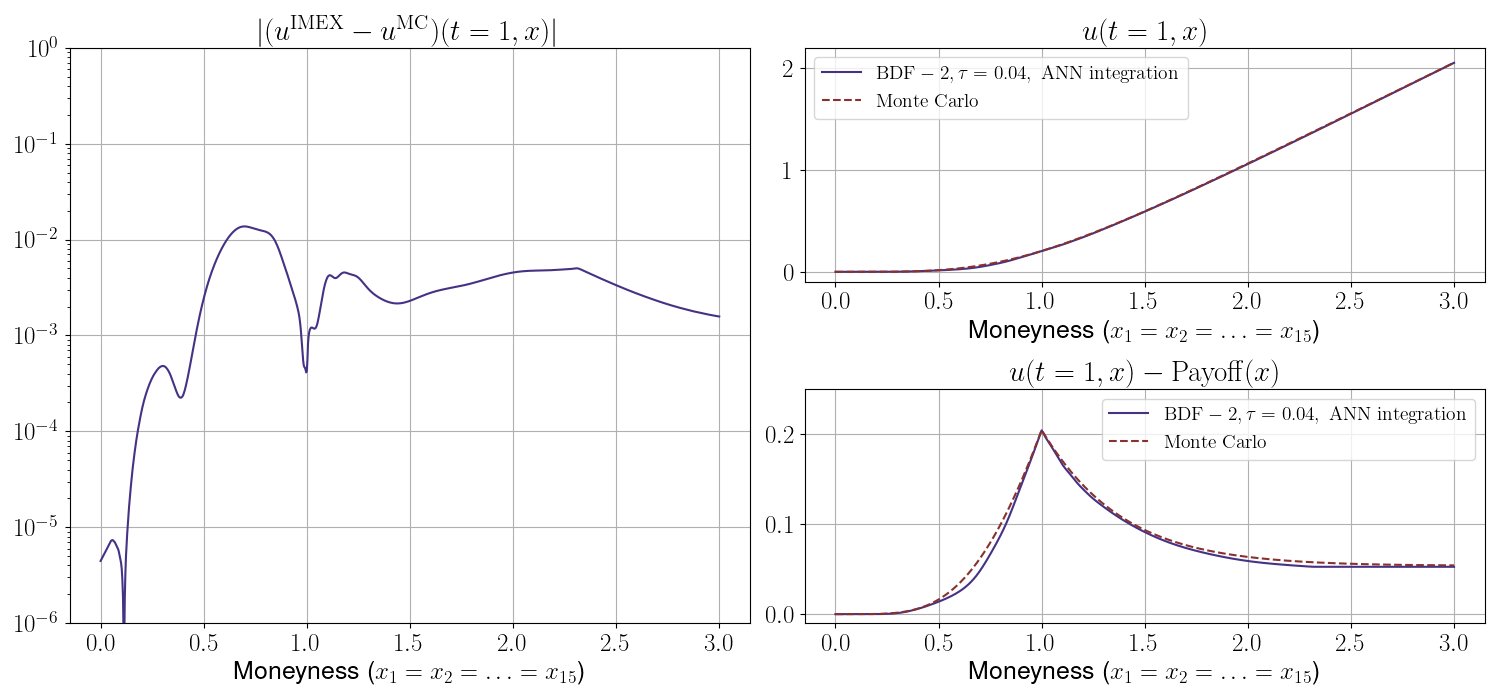}
    \caption{Basket option prices (top right), differences between price and payoff (bottom right) and differences between the proposed method and quasi Monte Carlo (left), using the BDF-2 scheme and an ANN for the calculation of the integral operator, for $d=15$.}\label{fig:15_bdf2_auto.png}
\end{center}
\end{figure}

We further test our numerical methodology in a rather challenging high-dimensional option pricing scenario.
In particular, we consider the case of a 15-asset European basket call option, as described earlier, with the maturity time being one year, \textit{i.e.} \(T=1\).
We employ time steps of \(\tau=0.01\) for the implicit-explicit Euler method and \(\tau = 0.03\) for the BDF-2 scheme, respectively.
Figure \ref{fig:15_euler_auto.png} presents the pricing results for the implicit-explicit Euler scheme, whereas Figure \ref{fig:15_bdf2_auto.png} provides the corresponding results for the BDF-2 scheme.
Both methods are performing very well once again, and the important observation is that the absolute error remains 
relative small, although the dimension of the problem is now three times higher.
The BDF-2 scheme is now 1.7 times faster than the implicit-explicit Euler method, however the time-step is now smaller compared to the case with $5$ assets.

\subsection{Hyperparameters and performance}

We report the hyperparameters, the computational setup, and the computational run times of the implementation.
The neural networks used consist of \(2\) DGM layers with a width of \(2^6\) neurons each.
The weights are initialized via the Xavier initialization.
The optimization is driven by the Adam optimizer using a fixed learning rate \(\alpha = 3\cdot 10^{-4}\).
During the initialization phase (\(t=0\)), the neural networks (for the solution and for integration where applicable) are trained for a total of \(2^{15}\) epochs, using  \(2^{15}\) sampled points at each epoch. 
A combination of uniform sampling for the moneyness and then sampling from the Dirichlet distribution is used during this step, providing a better initial start for the subsequent training steps.
For the first time step (\(t=\tau\)), we perform \(2^{14}\) and \(2^{15}\) epochs for the 5-asset and 15-asset cases, respectively.
Subsequently,  \(2^{12}\) (5-assets) and \(2^{13}\) (15-assets) epochs are performed for the following time steps.
At each iteration, a total of \(2^{12}d\) points are sampled in \(\mathbb R^d_+\). 
We have used scrambled Sobol sequences, \textit{cf.} \citet{SOBOL,owen1998}, for the sampling, which provide low-discrepancy quasi-random samples. The algorithms have been executed on a computational node of the DelftBlue supercomputer. 
In particular, we used 12 CPU cores (Intel XEON E5), 64GB RAM, and an NVIDIA Tesla V100S with 32GB RAM.

\subsection{Basic comparison with other ML solvers}

In an effort to position the method proposed above in terms of computational cost, as well as comment on its benefits, we now provide a basic comparison against two popular alternative methods: the deep BSDE solver with jumps by \citet{Gnoatto2022} and the deep Galerkin method by \citet{Sirignano2018}; in the latter method, we incorporate the ANN approximation of the integral operator implicitly.

For the Deep BSDE solver, we adapt the code provided by \citet{Gnoatto2022} on GitHub\footnote{\url{https://github.com/AlessandroGnoatto/DeepBsdeSolverWithJumps}}, which prices European basket call options under a jump-diffusion model with uncorrelated assets. To ensure fair comparison, we also consider uncorrelated assets for the other two methods.

We run the pricing schemes for a European basket call option with \(5\) underlying assets, using the parameters:
\(T=1,\  K= 1,\  \sigma_i = 0.5,\ r = 0.05,\  \lambda = 1,\  \mu_{J_i} = 0, \ \sigma_{Ji}=0.5\). 
For the deep BSDE solver, we solve for the initial moneyness value of \(x_i = 1\), while in the other two methods it is possible to price the option for an array of initial values at once.

Due to library incompatibilities with NVIDIA drivers, we have been unable to execute a CUDA-accelerated version of the deep BSDE solver. 
To facilitate a fair computational time assessment we, thus, compare CPU runtimes for all methodologies. 
We expect that CUDA execution would likely result in speedups of 10 times or more for all methods in a similar fashion. 
Moreover, to ensure a robust Monte Carlo (MC) estimation, we run 10 million iterations, leading to a standard deviation of \(\mathrm{std}=1.147\cdot 10^{-4}\). Thus, we feel it is appropriate to consider the Monte Carlo solution as the ground truth.

Before providing the results of the basic comparison, we highlight the key differences on how these methods approach the solution space. 
The MC method solves for a single time-space point, providing an estimate for a specific set of initial conditions at maturity. 
The deep BSDE solver provides solutions across multiple time steps but for fixed spatial values, \textit{i.e.}, fixed initial asset or moneyness values. 
In contrast, both DGM and the deep IMEX scheme, proposed in this work, provide solutions that can be evaluated across the entirety of a space-time domain.
This broader coverage makes DGM and deep IMEX more versatile but, as expected, they are also more computationally intensive. For instance, a key advantage of being provided with a complete solutions across the space-time domain, allows for straightforward calculation of Greeks. The latter is more challenging in local solvers, such as MC or deep BSDE.

We note that both deep IMEX and DGM are implemented with the same network architecture given in Section \ref{net_arch}. To ensure, however, comparable accuracy, we test two different configurations for the DGM method: one with an additional layer
and another with two additional layers, compared to the deep IMEX architecture. 
The results, including option values, absolute errors against the ``ground truth'' provided by MC, and runtimes, are summarized in Table~\ref{table:comparison}. 
This basic comparison shows that the proposed deep IMEX approach is competitive with both MC and deep BSDE methodologies in terms of error and runtime, (less than twice as expensive as MC and less then three times as expensive as deep BSDE, for similar accuracy) to produce a complete space-time solution. Moreover, deep IMEX appears to be more economical than DGM in terms of runtime and parameter cardinality.

\begin{table}[th]
\begin{center}
    \begin{tabular}{r|ccccc}
                  & MC & Deep IMEX & Deep BSDE & DGM (+1 layer)  & DGM (+2 layers) \\
                  \hline 
Value             & \(0.1647\)   & \(0.1633\)    & \(0.1661\)    & \(0.1509\) & \(0.1669\)   \\
Abs. Error        & -            & \(1.41E{-3}\) & \(1.36E{-3}\)    & \(1.38E{-2}\) & \(2.14E{-3}\)   \\
Runtime (min) & \(533\)      & \(885\)  & \(376\)       & \(1330\)  & \(1634\)       
\end{tabular}
\caption{\label{table:comparison}Comparison of pricing methods for a European basket call option of 5 assets under a jump-diffusion model.}
\end{center}
\end{table}



%% file: conclusions.tex
This work develops a novel deep learning method for the pricing of European basket options in models that follow jump-diffusion processes.
To address the intricacies of the problem, we introduce a decomposition technique, expressing the option price as the sum of an unknown component (time value) and a known lower-bound function (intrinsic value). 
The incorporation of a domain truncation method further enhances the accuracy of our numerical schemes. 
By projecting option prices onto a bounded subset and efficiently approximating solutions for extreme moneyness values within this truncated domain, we arrive at an accurate and reliable approximation of the underlying solution to the PIDE problem.
The combination of the deep implicit-explicit minimizing movement methodology, the decomposition of the solution, and the domain truncation method form a robust toolkit for enhancing the accuracy and efficiency of option pricing methods in advanced financial models, by providing a complete solution at every point of the space-time domain. 
As such, it is straightforward to infer further quantities, such as Greeks, from the computed space-time solution. A basic comparison with popular and successful approaches further showcases the practical relevance of the proposed method.





